\documentclass[a4paper, aip, jap, reprint, superscriptaddress]{revtex4-1}

\usepackage{graphicx}

\newcommand{\mum}{\mathrm{\mu m}}
\newcommand{\mus}{\mathrm{\mu s}}

\begin{document}

\title{High optical efficiency and photon noise limited sensitivity of microwave kinetic inductance detectors using phase readout.}

\author{R.M.J. Janssen}
\email{r.m.j.janssen@tudelft.nl}
\affiliation{Kavli Institute of Nanoscience, Faculty of Applied Sciences, Delft University of Technology, Lorentzweg 1, 2628CJ Delft, The Netherlands}

\author{J.J.A. Baselmans}
\affiliation{SRON Netherlands Institute for Space Research, Sorbonnelaan 2, 3584CA Utrecht, The Netherlands}

\author{A. Endo}
\affiliation{Kavli Institute of Nanoscience, Faculty of Applied Sciences, Delft University of Technology, Lorentzweg 1, 2628CJ Delft, The Netherlands}

\author{L. Ferrari}
\affiliation{SRON Netherlands Institute for Space Research, Landleven 12, 9747AD Groningen, The Netherlands}

\author{S.J.C. Yates}
\affiliation{SRON Netherlands Institute for Space Research, Landleven 12, 9747AD Groningen, The Netherlands}

\author{A.M. Baryshev}
\affiliation{SRON Netherlands Institute for Space Research, Landleven 12, 9747AD Groningen, The Netherlands}
\affiliation{Kapteyn Astronomical Institute, University of Groningen, P.O. Box 800, 9700 AV Groningen, The Netherlands}

\author{T.M. Klapwijk}
\affiliation{Kavli Institute of Nanoscience, Faculty of Applied Sciences, Delft University of Technology, Lorentzweg 1, 2628CJ Delft, The Netherlands}
\affiliation{Physics Department, Moscow State Pedagogical University, Moscow, 119991, Russia}

\begin{abstract}
We demonstrate photon noise limited performance in both phase and amplitude readout in microwave kinetic inductance detectors (MKIDs) consisting of NbTiN and Al, down to 100 fW of optical power. We simulate the far field beam pattern of the lens-antenna system used to couple radiation into the MKID and derive an aperture efficiency of 75\%. This is close to the theoretical maximum of 80\% for a single-moded detector. The beam patterns are verified by a detailed analysis of the optical coupling within our measurement setup.
\end{abstract}

\maketitle

\begin{figure}
\centering
\includegraphics[width=1.0\columnwidth]{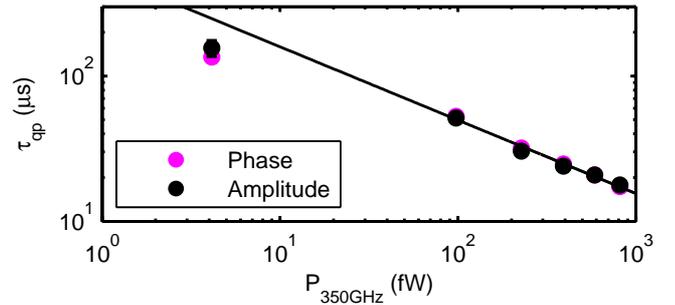}
\caption{The quasiparticle lifetime, determined from the phase (magenta) and amplitude (black) noise roll-off, as a function of optical power absorbed by the MKID, $P_{350GHz}$. The Lorentzian fit gives a typically error of 5\% in the estimated lifetime. As a result, errorbars are usually smaller than the symbol size. A fit to the measured lifetimes for $P_{350GHz} > 100$ fW shows that $\tau_{qp} \propto P_{350GHz}^{-0.50\pm0.02}$.}
\label{Figure:Lifetime}
\end{figure}

\begin{figure}
\centering
\includegraphics[width=1.0\linewidth]{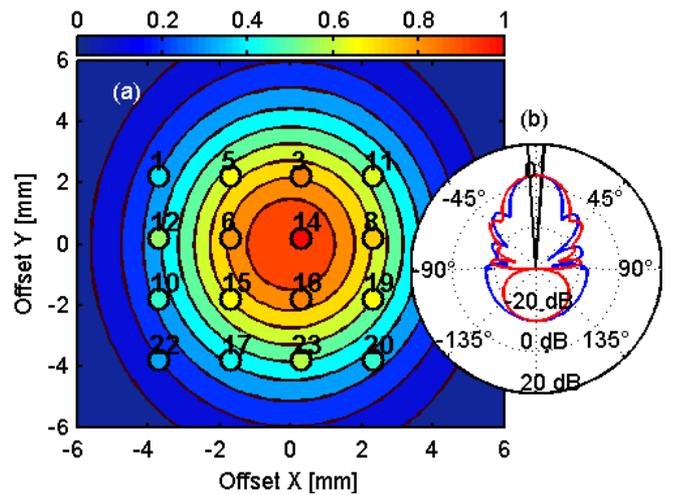}
\caption{(a) This contour plot shows the power received by the hybrid MKIDs as a function of the offset between the lens and aperture center. The scale is normalized to the power received at co-alignment. Overlaid in black circles are the positions of the 16 MKIDs with lenses. The color inside each circle indicates the relative change in resonance frequency between 5 and 90 fW of optical loading, which is a proxy for the relative absorbed power. (b) The gain (in dB) of the complete lens-antenna system as a function of angle (in degrees) as obtained from CST. The red and blue line are the two orthogonal cross-sections. The black solid lines show the opening angle of the aperture used in this experiment.} 
\label{Figure:ArrayResponse}
\end{figure}

In the next decades millimeter and sub-mm astronomy require\cite{Scott2010} large format imaging arrays to complement the high spatial resolution of the Atacama Large Millimeter/submillimeter Array\cite{Brown2004}. The desired sensors should have a background limited sensitivity and a high optical efficiency and enable arrays of up to megapixels in size. The most promising candidate to fulfill these requirements are microwave kinetic inductance detectors (MKIDs)\cite{Day2003} due to their inherent potential for frequency domain multiplexing. MKIDs are superconducting resonators, thousands of which can be coupled to a single feedline. Each resonator is sensitive to changes in the Cooper pair density induced by absorption of sub-mm radiation. By monitoring the change in either phase or amplitude of the complex feedline transmission at the MKID resonance one can measure the absorbed photon power. Using amplitude readout photon noise limited performance has been shown\cite{Yates2011}. However, for practical applications two key properties need to be demonstrated: (1) Photon noise limited operation in phase readout. (2) A measurement of the aperture efficiency\cite{Rohlfs2004}, which describes the absolute optical coupling of a MKID imaging array to a plane wave.\\
In this letter we present antenna coupled hybrid NbTiN-Al MKIDs designed for groundbased sub-mm astronomy. We show that these devices achieve photon noise limited performance in both amplitude and phase readout. Through a detailed analysis of the optical coupling within our setup we validate the simulation of the lens-antenna far field beam pattern. From this we derive an aperture efficiency of 75\%. This is close to the theoretical maximum of 80\% for a single-moded detector.
\begin{figure}[b]
\centering
\includegraphics[width=1.0\columnwidth]{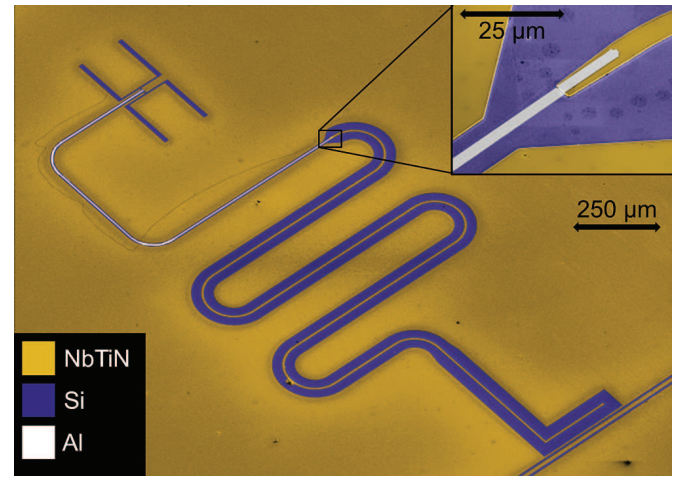}
\caption{Scanning electron micrograph of the antenna-coupled hybrid NbTiN-Al MKIDs used. A wide NbTiN CPW resonator is used to minimize the two-level system noise contribution. At the shorted end, where the planar antenna is located, one millimiter of CPW is reduced in width and the central line is made from thin Al. The Al is galvanically connected to the NbTiN at both ends (inset).}
\label{Figure:Device}
\end{figure}
\begin{figure*}[t]
\includegraphics[width=1.0\textwidth]{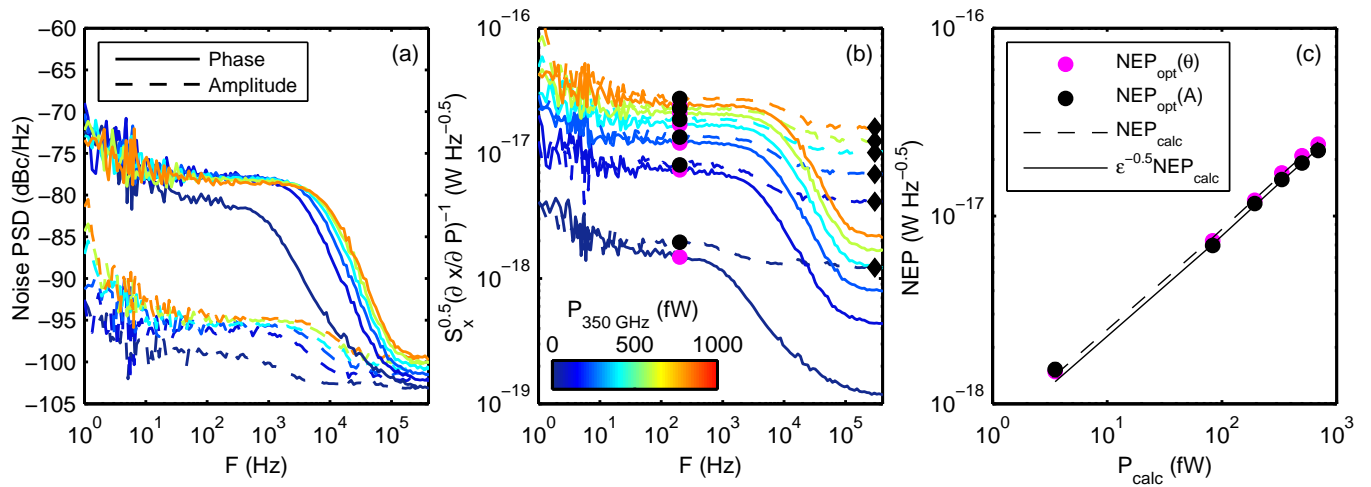}
\caption{(a) The power spectral density of the amplitude (dashed) and phase (solid) noise measured under various optical loading. A white noise spectrum is observed, the level of which is constant with loading power, for $P_{350GHz} > 100$ fW. The roll-off above 1 kHz is due to the quasiparticle lifetime, which is reduced by an increasing optical load. Note that the photon noise level is 16 dB higher in phase readout. Nevertheless, at a given loading the NEP is the same for phase and amplitude readout as shown in (b). (b) The optical NEP\cite{deVisser2013,TimeFactor} for amplitude (dashed) and phase (solid) readout as a function of optical loading. (c) The optical NEP measured at a modulation frequency of 200 Hz, $NEP_{opt}$, as a function of the estimated optical loading, $P_{calc}$. For every optical loading the NEP in phase (magenta dots) and amplitude (black dots) readout is equal, if for the latter the amplifier contribution (black diamonds) is subtracted. This contribution is insignificant in phase readout. The errorbars on the NEP measurement are of the order of the symbol size. The measured NEP follows the same slope as $NEP_{calc} \propto \sqrt{P_{calc}}$ (dashed black line). By fitting the relation between measured amplitude NEP and $NEP_{calc}$ (solid black line) $\epsilon=1.06\pm0.06$ is determined.}
\label{Figure:Spectra}
\end{figure*}\\
The device design, shown in Fig. \ref{Figure:Device}, aims to simultaneously maximize the phase response and minimize the two-level system (TLS) noise contribution\cite{Gao2007}. The device is a $L\approx5$ mm long quarter wavelength coplanar waveguide (CPW) resonator consisting of two sections. The first section ($\sim 4$ mm), at the open end of the resonator, is a wide CPW made entirely from 200 nm thick NbTiN. NbTiN has 10 dB lower TLS noise than conventional superconductors such as Al\cite{Barends2009}. The TLS noise is further reduced by the width of the CPW\cite{Barends2009}, 23.7 $\mum$ and 5.4 $\mum$ for the CPW gap and central line, respectively.\\
The second section (1 mm), at the shorted end of the resonator, is a narrow CPW with NbTiN groundplanes and a 48 nm thick Al central line. The Al is galvanically connected to the NbTiN central line (Fig. \ref{Figure:Device} inset) and the NbTiN groundplane at the resonator short. The NbTiN is lossless for frequencies up to the gap $2\Delta_0/h = 1.1$ THz ($T_c \approx 14$ K). Any radiation with a frequency $0.09 < \nu < 1.1$ THz is therefore absorbed in the Al ($T_c = 1.28$ K) central line of the second section. The optically excited quasiparticles are trapped in the Al, because it is connected to a high gap superconductor. This quantum well structure confines the quasiparticles in the most responsive part of the MKID and allows us to maximize the response by minimizing this active volume. Therefore, we use a narrow CPW in section two, 2.3 $\mum$ and 3.7 $\mum$ for the central line and slots, respectively. Using a narrow Al line at the shorted end of the MKID does not increase the TLS noise significantly, because of the negligible electric field strength in this part of the detector.\\
At the shorted end of the resonator light is coupled into the device through a single polarization twin-slot antenna, which is optimized for $\nu=350$ GHz. The advantage of using antenna coupling is that it can be designed independently from the distributed CPW resonator. The disadvantage is that the antenna occupies only $\sim 1\%$ of the total pixel footprint. To achieve a high filling fraction we use elliptical lenses to focus the light on the antennas.\\
The design presented here is an improvement on that by Yates et.al.\cite{Yates2011}. Our design has a wider body in section one, which provides $\sim 7$ dB reduction of the TLS noise. In addition we have thinner Al, 48nm instead of 80 nm. This increases the kinetic inductance fraction by 45\% to $\alpha=0.09$ and reduces the volume by 40\%. Both give a linear increase in the phase response.\\
An array of 24 pixels has been fabricated\cite{Lankwarden2012,Airbridges}  on a high resistivity ($>10$ $\mathrm{k\Omega \ cm}$) $\langle100\rangle$-oriented Si substrate. All pixels are capacitively coupled to a single feedline with a coupling $Q_c\sim 58\mathrm{k}$, which is matched to the $Q_i$ expected for an optical loading of $\sim 10$ pW. After mounting an array of 16 laser machined Si lenses with a diameter of 2 mm on the central pixels, the array is evaluated using a pulse tube pre-cooled adiabatic demagnetization refrigerator with a box-in-a-box cold stage design\cite{Baselmans2012b}. In this design the array is fully enclosed in a 100 mK environment with the exception of a 2 mm aperture, which is located 15.05 mm above the approximate center of the MKID array. This aperture is isotropically illuminated by a large temperature-controlled blackbody\cite{deVisser2013}. Two metal mesh filters provide a minimum rejection of 20 dB at all wavelengths outside the 50 GHz bandpass centered on $\nu=350$ GHz. This allows us to create a variable unpolarized illumination over a wide range of powers.\\
Fig. \ref{Figure:Spectra}(a) shows the amplitude and phase noise spectra measured for a typical device as a function of the optical power absorbed in the Al, $P_{350GHz}$. In this figure we observe two characteristics that prove our device is photon noise limited:
\begin{enumerate}
\item The noise spectra in both phase and amplitude are white with a roll-off given by the quasiparticle lifetime, $\tau_{qp}$, or resonator ring time, $\tau_{res}$. For our devices we observe a white noise spectrum for $P_{350GHz} \geq 100$ fW, which has a roll-off due to $\tau_{qp}$, because $\tau_{qp}>\tau_{res}$.
\item When reducing the optical loading from the photon noise limited situation one should observe at negligible power levels the transition to a noise spectrum that is limited by intrinsic noise sources of the detector. At a negligible optical loading ($P_{350GHz} = 4$ fW) the phase noise spectrum is no longer white and the noise level in both phase and amplitude readout is lower with respect to $P_{350GHz}>4$ fW.
\end{enumerate}
Fig. \ref{Figure:Spectra} shows three more features, which may be present in a photon noise limited MKID.\\
The photon noise level of the spectra observed in Fig. \ref{Figure:Spectra}(a) is independent of the optical loading, because the product of quasiparticle number and quasiparticle lifetime is constant\cite{Baselmans2012a,deVisser2013} and the loaded Q did not change.\\
By fitting a lorentzian roll-off\cite{deVisser2013} to the spectra presented in Fig. \ref{Figure:Spectra}(a) we derive the quasiparticle lifetime as a function of the optical loading. Fig. \ref{Figure:Lifetime} shows that the quasiparticle lifetime obtained from phase and amplitude readout is equal for all loading levels. For $P_{350GHz} > 100$ fW the quasiparticle lifetimes show a $\tau_{qp} \propto P_{350GHz}^{-0.50\pm0.02}$ relation, which matches the expected $\tau_{qp} \propto 1/\sqrt{P_{350GHz}}$ relation\cite{Baselmans2012a,BCS1957,Kaplan1976} for a homogeneously illuminated superconductor. At $P_{350GHz} = 4$ fW we observe $\tau_{qp} = 150$ $\mus$, which deviates from the trend set by the photon noise limited regime and is significantly lower than the $\tau_{qp} \approx 2$ ms observed by de Visser et.al.\cite{deVisser2011} in similar Al. Measurements on hybrid NbTiN-Al MKIDs with a varying length of the Al section show that the quasiparticle lifetime increases with the Al length. Based on this we tentatively conclude that the reduced lifetime we observe is due to poisoning by quasiparticles entering from the NbTiN.\\
Fig. \ref{Figure:Spectra}(b) and \ref{Figure:Spectra}(c) show the observed Noise Equivalent Power (NEP)\cite{deVisser2013,TimeFactor}, the level of which only depends on the optical loading and is thus equal in amplitude and phase readout. The NEP follows the $NEP \propto \sqrt{P_{350GHz}}$ relation expected for photon noise limited MKIDs\cite{Yates2011,Baselmans2012a,NEPscale}.\\
Given the equal NEP values, Fig. \ref{Figure:Spectra}(a) shows why phase readout is preferred for practical background limited systems. The photon noise level in phase readout is 16 dB higher than that in amplitude readout, thereby relaxing the dynamic range requirements of the readout electronics. We estimate\cite{Kester2005,Boyd1986} that with a state-of-the-art readout system based upon the E2V EV10AQ190 analog-to-digital converter (ADC) we can simultaneously read out approximately 1800 of the presented NbTiN-Al MKIDs in phase readout if we accept a 10\% degradation of observing time due to the noise added by the ADC alone. Using the same electronics amplitude readout would only allow 30 pixels.
\\
In the photon noise limited regime it is favorable to have a high optical or aperture efficiency\cite{Rohlfs2004}, $\eta_A$, because the observation time required to achieve a given signal-to-noise follows $t_{\sigma} \propto \eta_A^{-1}$. The most reliable way to determine the aperture efficiency of a MKID is through the measurement of the photon noise. Yates et.al.\cite{Yates2011} used this approach and determined the aperture efficiency by comparing the measured NEP to the NEP expected for a perfect absorber with the same area as a single pixel. The latter was determined using the geometrical throughput between the detector and the illuminating aperture. However, this approach is only valid if the gain of the antenna is equal to its maximum value within the entire angle spanned by the aperture.\\
We determine the aperture efficiency from the full far field beam pattern, which is obtained from a simulation of the complete lens-antenna system using CST Microwave Studio. This gives us the freedom to adjust the design as required and allows a calculation of $\eta_A$ independent of the measurement setup. However, the simulated beam pattern does require experimental verification. We will first show how we verify the coupling efficiency, $C_{\nu}$, which describes the reflection losses due to mismatches between the antenna and the resonator CPW, and the gain pattern\cite{Gnorm}, $G_{\nu}(\Omega)$, as a function of the angular direction, $\Omega$, from the microlens focus. The gain pattern of our lens-antenna system is shown in Fig. \ref{Figure:ArrayResponse}(b). After the verification we will use $C_{\nu}$ and $G_{\nu}(\Omega)$, which we obtain from CST Microwave Studio, to calculate $\eta_A$.\\
We verify $C_{\nu}$ and $G_{\nu}(\Omega)$ using the photon noise limited optical NEP, $NEP_{opt}$, as the experimental observable. We compare $NEP_{opt}$ to the photon noise limited NEP we expect, $NEP_{calc}$, from the resonator. To calculate $NEP_{calc}$ we need to know the power coupled to the lens-antenna system, $P_{calc}$, which is given by
\begin{equation}
P_{calc} = \frac{c^2}{4\pi} \int_{\nu} \int_{\Omega\in A_{ap}} \frac{F_{\nu}B_{\nu}(T_{BB})}{2\nu^2} \ C_{\nu}G_{\nu}(\Omega) \ d\Omega d\nu
%\label(Eq:Pest)
\end{equation}
Here $c$ is the speed of light, $F_{\nu}$ the filter transmission and $B_{\nu}(T_{BB})$ Planck's law for a blackbody temperature $T_{BB}$. The factor $1/2$ takes into account that we receive only a single polarization. The second integral evaluates the solid angle $\Omega$ over the aperture area, $A_{ap}$. The rest of the detector enclosure is a 100 mK absorber that has a negligible emission at 350 GHz. Included in this calculation is the experimentally measured lateral shift between each pixel and the aperture. The effect of lateral deviations from co-alignment are shown in Fig. \ref{Figure:ArrayResponse}(a). The contours in this figure show the predicted reduction in received power as a function of the lateral translation between the lenses' optical axis and aperture center. The contours are normalized to a co-aligned system. The circles in Fig. \ref{Figure:ArrayResponse} indicate the positions of the 16 lensed pixels. The color indicates the relative frequency change between 5 and 90 fW of optical loading, which roughly approximates the relative absorbed power for all 16 pixels. The qualitative match is striking and assures us of the shape of $G_{\nu}(\Omega)$. \\
We can now define the power error ratio $\epsilon$ as the discrepancy between the calculated and measured photon noise limited NEP.
\begin{eqnarray}
\epsilon &=& \frac{NEP^2_{calc}}{NEP^2_{opt}} \nonumber \\ 
&=& \frac{2P_{calc}(h\nu(1+mB)+\Delta/\eta_{pb})}{NEP^2_{200Hz}-NEP^2_{det}}
%\label(Eq:Epsilon)
\end{eqnarray}
Here $h\nu$ is the photon energy of the incoming radiation, $(1+mB)$ the correction to Poisson statistics due to wave bunching\cite{deVisser2013}, $\Delta$ the superconducting energy gap of the absorbing material and $\eta_{pb}= 0.57$ the pair breaking efficiency inside this material. $NEP_{opt}$ is equal to the measured NEP at a modulation frequency of 200 Hz, $NEP_{200Hz}$, after correction for any detector noise contribution to the NEP, $NEP_{det}$. In amplitude readout this is the amplifier noise contribution, which we estimate from the NEP value at a modulation frequency of 300 kHz (black diamonds in Fig. \ref{Figure:Spectra}(b)). In phase readout $NEP_{det}=0$ as both the frequency independent amplifier noise contribution, observed above the roll-off frequency, and the $1/\sqrt{F}$ TLS noise contribution, observed below 10 Hz, are insignificant at 200 Hz. We expect $\epsilon=1$, if the description of the optical power flow is complete and the simulated beam patterns are correct.
Fig. \ref{Figure:Spectra}(c) shows the measured optical NEP, $NEP_{opt}$, as a function of $P_{calc}$ for phase (magenta dots) and amplitude (black dots) readout. The measured photon noise NEPs from phase and amplitude are within $2\sigma$ of each other at all loading levels. The solid black line shows the best fit for the expected linear relation, $NEP_{opt} = NEP_{calc}/\sqrt{\epsilon}$. For the presented MKID, numbered 16, $\epsilon=1.06\pm0.06$, if we disregard the lowest loading. For a different MKID, numbered 3, the above analysis yields $\epsilon = 1.09\pm0.13$.\\
From the verified far field beam pattern we can determine $\eta_A$, which is mathematically defined as
\begin{equation}
\eta_A = \frac{A_e}{A} = \frac{\lambda^2G_{\nu}(\Omega_0)C_{\nu}}{4 \pi A}
%\label(Eq:Pest)
\end{equation}
Here $A$ is the physical area covered by the pixel; $\lambda$ and $\nu$ are the wavelength and frequency of the observed radiation, respectively; and $\Omega_{0}$ is the direction of the maximum gain. Using the circular area of the lenses $A=\pi$ $\mathrm{mm^2}$, $C_{350GHz}=0.98$ and the gain of the CST beam pattern at broadside, $G_{\nu}(\Omega_0) = 5.0$ dB, an aperture efficiency of 75\% is determined for a single pixel. The maximum achievable aperture efficiency of a circular antenna illuminated by a single-moded gaussian beam is\cite{QuasiOptics} $\eta_A=0.80$. For the measured array the filling fraction of the square packing means we have a total array aperture efficiency of 57\%. Using an array with hexagonal packing the total array aperture efficiency can be increased to 66\%.\\
In conclusion, we present hybrid NbTiN-Al MKIDs, which are photon noise limited in both phase and amplitude readout for loading levels $P_{350GHz} \geq 100$ fW with an aperture efficiency of 75\%. The photon noise level will allow us to simulatenously read out approximately 1800 pixels using state-of-the-art electronics to monitor the phase. Given these specifications, hybrid NbTiN-Al MKIDs should enable astronomically usable kilopixel arrays for sub-mm imaging and moderate resolution spectroscopy.

\begin{acknowledgments}The authors thank T. Zijlstra, D.J. Thoen and Y.J.Y. Lankwarden for sample fabrication and P.J. de Visser for helpful discussion. T.M. Klapwijk and R.M.J. Janssen are grateful for support from NOVA, the Netherlands Research School for Astronomy, to enable this project. A. Endo is grateful for the financial support by NWO (Veni grant 639.041.023) and the JSPS Fellowship for Research Abroad. The work was in part supported by ERC starting grant ERC-2009-StG Grant 240602 TFPA (A.M. Baryshev). T.M. Klapwijk acknowledges financial support from the Ministry of Science and Education of Russia under contract No. 14.B25.31.0007.
\end{acknowledgments}


\begin{thebibliography}{21}
\bibitem [{Tim()}]{TimeFactor}%
  \BibitemOpen
  \href@noop {} {\bibinfo  {journal} {For clarity the factor $\sqrt{(1+(2\pi
  F)^2\tau_{qp}^2)(1+(2\pi F)^2\tau_{res}^2)}$ is not included}\ }\BibitemShut
  {NoStop}%
\bibitem [{Air()}]{Airbridges}%
  \BibitemOpen
  \href@noop {} {\bibinfo  {journal} {Measurements have shown that the amount
  of direct absorption by the resonator is negligible when lenses are used.
  Therefore, we do not use airbridges to connect the groundplanes of the CPW
  resonator}\ }\BibitemShut {NoStop}%
\bibitem [{NEP()}]{NEPscale}%
  \BibitemOpen
  \href@noop {} {\bibinfo  {journal} {The $NEP \propto \sqrt{P_{350GHz}}$
  expected for photon noise limited MKIDs would also be observed for amplifier
  limited MKIDs, due to the $\tau_{qp} \propto 1/\sqrt{P_{350GHz}}$ relation.
  However, in this situation the roll-off in the spectra of Fig.
  \ref{Figure:Spectra}(a) would not be observed}\ }\BibitemShut {NoStop}%
\bibitem [{Gno()}]{Gnorm}%
  \BibitemOpen
  \href@noop {} {\bibinfo  {journal} {We manually normalize the simulated beam
  pattern to obtain $G_{\nu}$}\ }\BibitemShut {NoStop}%
\end{thebibliography}
\end{document}